# Ultrafast demagnetization dynamics in Ni: role of electron correlations

Shree Ram Acharya,[1] Volodymyr Turkowski,[1,*] Guo-ping Zhang,[2] Talat S. Rahman[1]

[1]Department of Physics, University of Central Florida, Orlando, FL 32816, USA

[2] Department of Physics, Indiana State University, Terre Haute, IN 47809, USA

**Experimental observations of the ultrafast (less than 50 fs) demagnetization of Ni have so far defied theoretical explanations particularly since its spin-flipping time is much less than that resulting from spin-orbit and electron-lattice interactions. Through the application of an approach that benefits from spin-flip time-dependent density-functional theory and dynamical mean-field theory, we show that proper inclusion of electron correlations and memory (time-dependence of electron-electron interaction) effects leads to demagnetization at the femtosecond scale, in good agreement with experimental observations. Furthermore, our calculations reveal that this ultrafast demagnetization results mainly from spin-flip transitions from occupied to unoccupied orbitals implying a dynamical reduction of exchange splitting. These conclusions are found to be valid for a wide range of laser pulse amplitudes. They also pave the way for *ab initio* investigations of ultrafast charge and spin dynamics in a variety of quantum materials in which electron correlations may play a definitive role.**

Ultrafast tuning of the magnetization in transition-metal ferromagnets by short laser pulses has attracted worldwide attention because of possible applications in ultrafast data storage, switches, and spintronics, to name a few. The unusual physical effects accompanying such as a fast - femto-second (fs) – demagnetization, namely non-trivial dynamics of electrons, spins and lattice, and their interactions (e.g., orbital momentum transfer between the subsystems) have also challenged standard theoretical explanations (for an over-review, see ref.[1]). Beginning with the pioneering experimental work of Beaurepaire et al. on nickel, [2] which displays an ultrafast demagnetization when excited by an ultrashort laser pulse, the subject continues to be examined both experimentally and theoretically. A number of experimental observations [2-11] have now confirmed that this laser-induced demagnetization in bulk and thin film Ni takes place at the sub-picosecond time scale. The latest results showing the time scale to be about 20 fs [11]. Not surprisingly most theoretical studies have attempted to trace the origin of this demagnetization to intricacies in the electronic and spin structure of the system. Probably the simplest is the phenomenological three-temperature (3T) model [2], in which magnetization dynamics is characterized by an effective spin temperature, which equilibrates through energy exchange between the spin subsystem and electron and phonon baths. This model can be used to fit experimentally measured electron and spin temperatures, but it does not provide a microscopic understanding of the processes involved in the demagnetization, except for the possible role of phonons which occurs at the post picosecond stage. A modified 3T model that adds electron-phonon momentum scattering events [12] and another that adds dynamical feedback exchange splitting between majority and minority bands [13] have also been proposed. Efforts have also been made to augment traditional spin wave theory with laser-induced spin-orbit torque[14] to explain the recent observations of ultrafast critical behavior in Ni [10].

Since phonons may not play a large role at fs, Zhang et al. [15] analyzed a model Hubbard Hamiltonian that related the demagnetization to a combined effect of the external laser field and spin-orbit coupling in the system, a conclusion later corroborated experimentally [9]. These Hubbard-model based studies (see Ref. [15] and references therein) have also aimed at understanding the role of electron correlations in the demagnetization but the interaction parameters (fitted to spectroscopic data) used in these studies are quite different from those that provide good agreement with experimental data on the Ni Curie temperature and high temperature magnetic moment [16]. Informative as these studies have been, a microscopic theory that explains ultrafast demagnetization of Ni is lacking.

*Ab initio* approaches ranging from rigid band DFT [17] to time-dependent spin-density functional theory (TD spin DFT) [18,19] have also had some success. A TDDFT study [19] based on the non-collinear local spin density exchange-correlation (XC) potential with no memory dependence did demonstrate a large decrease of the magnetic moment (~43%) in Ni within a few dozen fs, as a result of spin-orbit interactions between the excited and ground state electrons. However, the pulse fluence used [19] were two order of magnitude larger and the pulse wavelength one-half those in experiments. On the other hand, incorporation of time dependent Liouville equation into DFT [18] with rescaled spin-polarized local density approximation (LDA) and a time-dependent attenuation factor found 10% decrease of magnetic moment for experimentally relevant pulse parameters [18]. Although the demagnetization is much less than the experimental value, this study [18] points to the importance of memory effects in XC potentials. To isolate the source of ultrafast demagnetization and expose further the relevance of memory effect, we present here a theoretical model in which we incorporate non-Markovian dynamics in non-collinear spin-density TDDFT [20,21] with electron correlation accounted in the XC kernel derived from dynamical mean field theory (DMFT) [22,23], which in turn is based on a Hubbard model suitable for transition metals with partially-filled d-orbitals. Apart from inherent inclusion of memory effects [24,25], the approach tracks electron correlations at time scale during which lattice effects may be neglected: the first (0-20 fs) of ultrafast spin dynamics in Ni.

Note that such a non-collinear theory would allow spins to flip (Fig.1) without requiring conservation of $S_z$, since the magnetization direction varies in space and the z-component of spin is not a good (global) quantum number. Realization of such systems could be the helical spin-wave ground state for γ-Fe [26] and systems with varying surface magnetization. One could also visualize a scenario in which the ground state constitutes of collinear spins but coupling of the spin-up and spin-down densities either through an external transverse magnetic perturbation or a spatially-dependent effective magnetic field generated by reorientation of the magnetic moments leads to non-collinearity.

In Fig. 2, we show the TDDFT results for the time dependence of demagnetization in Ni after laser pulse perturbation using three different XC kernels: 1) with full memory effects; 2) no memory effects; 3) memory effects in only the spin flip part. The demagnetization of 56% obtained with the XC with full memory effects is in a good agreement with what we can extract from experimental data, ~40% for pulse with $\hbar\omega$ = 2 eV and 7 mJ/cm$^2$ fluence [2] and ~70 % for pulse of $\hbar\omega$ = 1.55 eV and fluence of 7.36 mJ/cm$^2$ (absorbed fluence of 2.56 mJ/cm$^2$) ,[11] while the XC kernel with no memory effects results in about 25.8% demagnetization. Even though this demagnetization is almost two times smaller than that obtained with memory effects, it is still much larger than that that have been obtained from standard TDDFT, pointing to the importance of electron correlations which is inherent in our XC kernel. Fig. 2 also shows demagnetization of 50.6%, when memory effects are confined to the spin flip part of the XC kernel, indicating that the major channel of demagnetization is spin-up to spin-down transitions. To probe further the origin of this enhanced ultrafast effect, we plot in Fig. 3 the time-dependence of the matrix elements of the DMFT XC kernel. Interestingly, the time dependence of the matrix elements is dramatic only at short times (< 0.1 fs). Furthermore, the time dependence dies out within ~1 fs, which is of order of electron-electron scattering time in correlated materials.

The inset of Fig. 2 summarizes the demagnetization that we obtain in the limiting cases of no $f_{xc}$ and standard TDLDA. Note that the maximum demagnetization obtained for these cases is about 0.28%, a value close to that obtained in other theoretical studies and far from the experimental results, as summarized in the Introduction.

Analyses of the orbital and spin-resolved excited charge dynamics shown in Fig.4 provide further insights that the population of the excited spin-down electrons is significantly higher than that of those with spin-up. This result is consistent with the higher density of unoccupied spin-down states near the $E_F$ in Fig. 5 in which we show the spin- and orbital-projected density of d-states of Ni obtained with DFT and DFT+DMFT approaches. Among all spin-down orbitals, the higher excited charge density in the $d_{xy}$, $d_{yz}$, and $d_{xz}$ orbitals as compared to the $d_{z2}$, $d_{x2-y2}$ orbitals is also consistent with their relatively higher DOS near the $E_F$ in Fig. 5. In Fig. 5, one can also see that as a result of local correlation and exchange effects, the DMFT density of states is shifted to slightly lower energy with respect to the Fermi energy ($E_F$) as compared to the DFT solution. The change in the magnetic moment per Ni atom is minor: 0.64 $\mu_B$ in DFT and 0.61 $\mu_B$ in DMFT, which is a bit closer to the experimental value of $0.57\mu_B$ [27]. The reduction of the magnetic moment in DMFT may be ascribed to small increase in the occupancy of spin-down orbitals. Since the imbalance in the orbital occupancy that contributes to magnetization mostly results from 3d orbitals and the density of states of the d-orbitals is significantly higher than that of the p-orbitals in the vicinity of $E_F$, especially in the valence band, we consider only the d-orbitals in the study.

In agreement with the experimental result, our calculations show the strong dependence of demagnetization on the pulse field amplitude, i.e. intensity or fluence [11,12,28] and the demagnetization time increase with amplitude [29] (see Fig. SI. 1 in the Supplementary Information). Examination of the change of the (m-) components and of the total z-projection of the angular momentum during the process of demagnetization (Supplementary Information, Section IV) also reveal that change is small (see Fig. SI. 2 in supplementary information), which confirms that spin-flip processes play dominant role in the demagnetization.

In short, we have demonstrated here that spin-flip processes caused by electron correlations vary from site to site in a non-collinear spin system are capable of producing a large demagnetization in Ni in the experimentally-observed ~50 fs. It is important to stress that the presented scenario is valid only in the non-collinear spin system case in which spin spatial orientations vary site to site on the lattice. These non-collinearities are always present on surfaces. Layer-dependent absorption of light and magnetic anisotropy are examples of possible sources of spin non-collinearity. The lattice (phonon) effects may also contribute to the magnetization dynamics in this non-collinear scenario, but we expect them to be important at longer times.[12]

In this work, we have provided a theoretical understanding of the experimentally observed large, ultrafast demagnetization of Ni by tracing it to spin-flip transitions resulting from electron-electron correlations that occur at the fs time scale after perturbation by a laser pulse. The failure of prior DFT based methods to explain this large demagnetization stems from their inability to include memory dependence in XC functional. On the basis of TDDFT+DMFT approach which incorporates time-resolved Coulomb interactions we have demonstrated that the change of the spin-down density of states of Ni at the Fermi level coupled with spin-flip transitions is capable of inducing a strong ultrafast demagnetization in Ni, which is in good agreement with experimental data. The same TDDFT+DMFT calculations without time-resolved electron-electron interactions lead to only one-half the observed demagnetization. There are some open questions that need to be answered in the framework of TDDFT+DMFT, and probably the most important of them is the non-linear response (when one needs to go beyond the XC kernel approximation) and the longer-time dynamics, when the phonons become important, which we are planning to address in the future.

**Acknowledgements** We thank Carsten Ulrich for many helpful discussions and acknowledge partial support from DOE Grant No. DE-FG02-07ER46354. We would like to acknowledge the computational resources provided by the STOKES facility at the University of Central Florida. GPZ was supported by US Department of Energy research grant No. DE-FG02-06ER46304. This paper is dedicated to the memory of Eric Beaurepaire and Jean Yves Bigot whom we (TSR, SRA, VT) did not know personally but whose seminal works and beautiful writings inspired our efforts.

**Author Contributions** S.R.A. performed the calculations. V.T. led the development of the formalism. T.S.R. guided the project. S.R.A led the writing of the manuscript. All authors contributed to the analysis and discussion of results and writing of the manuscript.

**Additional Information** The authors declare no competing financial interests. Supplementary information is available for this paper. Correspondence and requests for materials should be addresses to Volodymyr.Turkowski@ucf.edu.

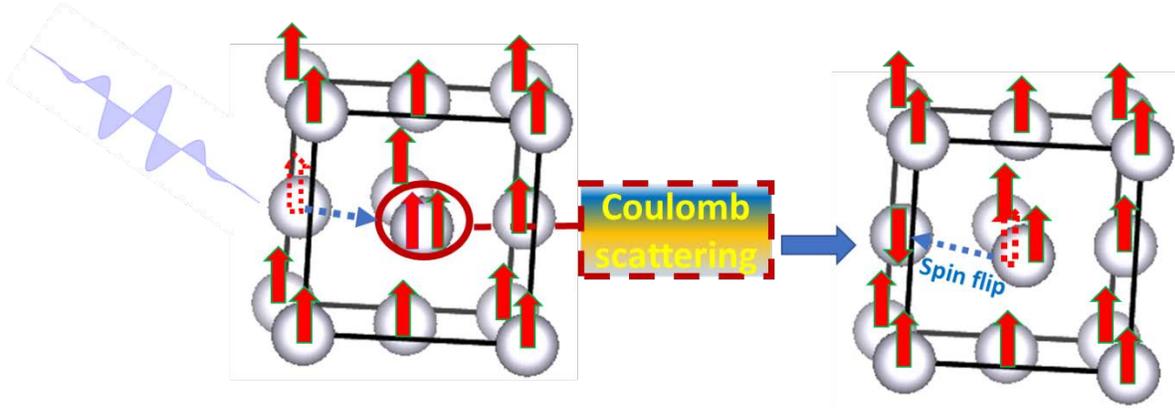

Fig. 1. Schematic representation of different stages of the laser pulse induced ultrafast demagnetization of Ni due to spin-flip processes.

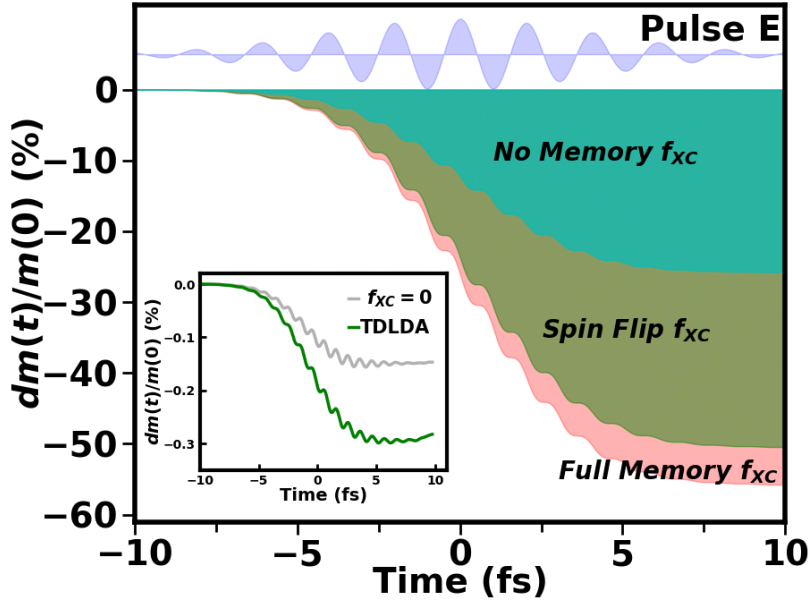

Fig. 2. The demagnetization dynamics, dm (t), calculated from eq. II.3 in SI by using XC kernel matrix with full memory dependence i.e., eq. (7) (edge of pink shaded area), with no memory dependence i.e., eq. (8) (edge of blue shaded area) and with only spin-up to spin-down flip part of the memory dependence i.e., keeping only $f_{XC32}$ in eq. (7) non-zero (edge of green shaded area). The amplitude, duration and frequency of pulse are taken as 0.05 V/Å, 7.2 fs and ħω = 2 eV,

respectively. The dynamics for limiting cases of no $f_{XC}$ and LDA XC in TDDFT is shown in inset plot.

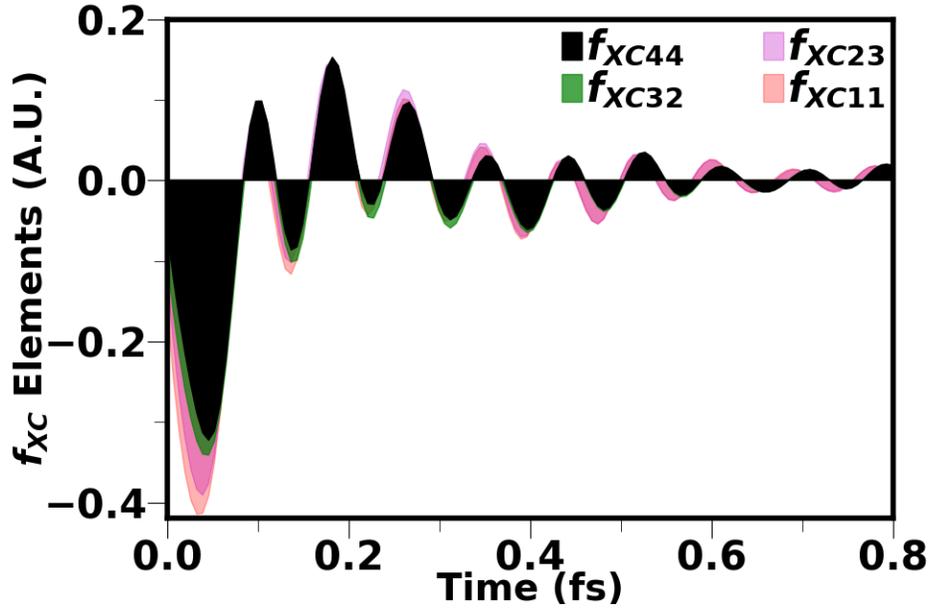

Fig. 3. Time-dependence of the non-zero components of the DMFT XC kernel of bulk Ni obtained from the Fourier transform of matrix elements in eq. (7).

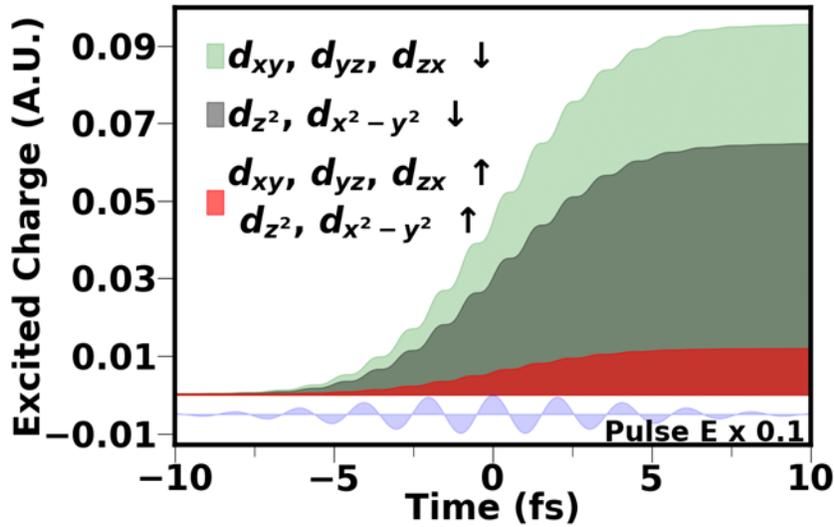

Fig. 4. Orbital and spin resolved dynamics of excited charge density obtained as the diagonal elements of density matrix.

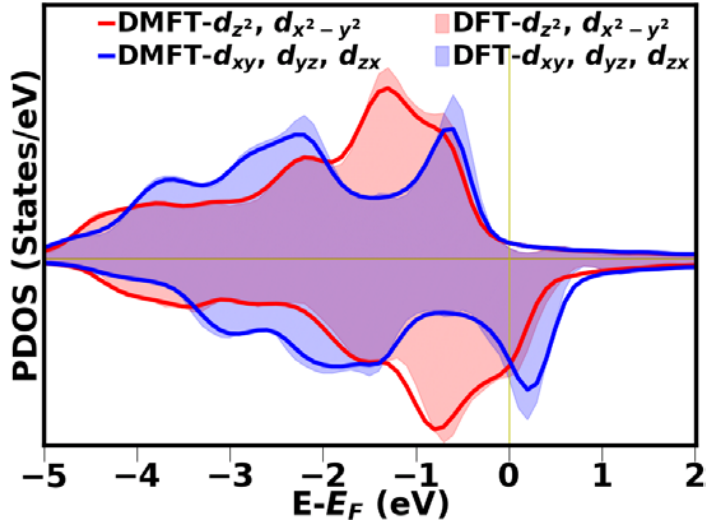

Fig. 5. The projected density of states (PDOS) of bulk Ni obtained using DFT (shaded area) and the spectral function obtained using DFT+DMFT approach (continuous curves) with on-site Coulomb interaction strength (U) = 3.0 eV and the strength of exchange interaction (J) = 0.9 eV. The PDOS of minority spin is flipped.

**METHODS**

As the first step, we perform a spin-polarized DFT [30,31] calculations using the Quantum Espresso package [32] to obtain the spin-resolved orbital density of states (DOS) and the corresponding Kohn-Sham eigenfunctions. The Perdew-Burke-Ernzerhof (PBE) functional [33] under the generalized gradient approximation (GGA) is used to describe the electronic exchange-correlation contribution to the total energy. The valence electron ($4s^2\ 3d^8$) wave functions are expanded using plane-wave basis up to a kinetic energy cut-off of 35 Rydberg (Ry) and the scalar-relativistic and ultra-soft pseudopotential is used to describe core electrons and the nuclei. The Brillouin zone of the unit cell of bulk Ni is represented by Monkhorst-Pack k-point scheme [34] with a 15×15×15 grid mesh. The structure is relaxed by allowing atomic position to change such that forces on an atom converge to $10^{-3}$ Ry/Å and the difference in total energy in successive electronic iterations convergence to $10^{-4}$ Ry. The calculated lattice constant of 3.52 Å is in agreement with experiments [35]. In the post-processing calculations of the DOS, 30×30×30 k-points and 20 bands are used.

For the effective Hubbard Hamiltonian solved within DMFT we choose the local Coulomb repulsion U = 3.0 eV and the exchange energy J = 0.9 eV obtained by constrained DFT [36], which successfully reproduced several experimental features of Ni [16,37]. To solve DMFT equations in the Matsubara (imaginary) frequency representation with discrete fermionic frequencies $i\omega_n = i\pi T(2n + 1)$, we take n = 250, $k_B T = 0.01$ eV and use the multi-orbital iterative perturbation theory (MO-IPT) approximation [38]. The Green's function obtained is then transformed into real frequency representation by using the Pade approximation [39]. With this set of values, we obtain the Curie temperature $T_c \approx 627$ K, in a rather good agreement with the experimental value of 631K [35].

To simulate electron dynamics of the system, we use spin-flip TDDFT Kohn-Sham equations for the spin wave functions, whose general form is:

$$\left[\left(-\frac{\nabla^2}{2m} + V_H[n](r,t)\right)\delta_{\sigma\sigma'} + V_{XC\sigma\sigma'}[n](r,t) + V_{ext\sigma\sigma'}(r,t)\right]\Psi_{k\sigma'}(r,t) = i\frac{\partial \Psi_{k\sigma}(r,t)}{\partial t}, \qquad (1)$$

where $-\frac{\nabla^2}{2m}$ is the kinetic energy operator, $V_H[n](r,t) = \int \frac{n(r',t)}{|r-r'|}dr'$ is the Hartree potential, $V_{XC\sigma\sigma'}[n](r,t)$ is the XC potential matrix, σ refers to spin indices, $V_{ext\sigma\sigma'}(r,t)$ is the external potential. The $V_{XC\sigma\sigma'}[n](r,t)$ is a functional of the spin-density matrix

$$n_{\sigma\sigma'}(r,t) = \sum_{k \leq k_F} \Psi_{k\sigma}(r,t) \Psi^*_{k\sigma'}(r,t). \qquad (2)$$

The $V_{ext\sigma\sigma'}(r,t)$ represents the laser pulse field and is taken as $\delta_{\sigma\sigma'} er \cdot E(t)$ using the dipole approximation which is valid situation in which pulse wave length is longer than the lattice parameters. The external electric field is taken as $E(t) = E_0 e^{-\frac{t^2}{\tau^2}} \cos(\omega t)$, where parameters $E_0$, $\tau$ and $\omega$ are the electric field amplitude, the pulse duration, and the field frequency, respectively. Unless specified otherwise, $E_0$ and $\hbar\omega$ used in this study are 0.05 V/Å and 2 eV, the same as in the TDLDA study [17], whereas $\tau$ is taken to be 7.2 fs, slightly less than that used in [17].

We use the linear response approximation in which the XC potential in eq. (1) can be expressed in terms of XC kernel matrix $f_{XC\sigma\sigma'\bar{\sigma}\bar{\sigma}'}(r,t,r',t')$ as

$$V_{XC\sigma\sigma'}[n](r,t) = V_{XC\sigma\sigma'}[n](r,t=0) + \sum_{\bar{\sigma},\bar{\sigma}'} \int f_{XC\sigma\sigma'\bar{\sigma}\bar{\sigma}'}(r,t,r',t') \, \delta n_{\bar{\sigma}\bar{\sigma}'}(r',t') \, dr'dt', \qquad (3)$$

where $V_{XC\sigma\sigma'}[n](r,t=0)$ is the static or DFT part of the XC potential and

$$f_{XC\sigma\sigma'\bar{\sigma}\bar{\sigma}'}(r,t,r',t') = \frac{\delta v_{XC\sigma\sigma'}[n](r,t)}{\delta n_{\bar{\sigma}\bar{\sigma}'}(r',t')}. \qquad (4)$$

In the DMFT approximation, $f_{XC\sigma\sigma'\bar{\sigma}\bar{\sigma}'}(r,t,r',t')$ becomes the product of the space- and time-dependent parts:

$$f_{XC\sigma\sigma'\bar{\sigma}\bar{\sigma}'}(r,r',t,t') = \delta(r-r') f^{DMFT}_{XC\sigma\sigma'\bar{\sigma}\bar{\sigma}'}(t-t'), \qquad (5)$$

where the time-dependent part $f^{DMFT}_{XC\sigma\sigma'\bar{\sigma}\bar{\sigma}'}(t-t')$ is obtained from Fourier transform of the frequency-dependent term $f^{DMFT}_{XC\sigma''\sigma''' \bar{\sigma}''\bar{\sigma}'''}(\omega)$ that satisfies the equation:

$$\chi_{\alpha\beta}(\omega) = \chi^{(0)}_{\alpha\beta}(\omega) + \sum_{\gamma,\delta} \chi^{(0)}_{\alpha\gamma}(\omega) f^{DMFT}_{XC\gamma\delta}(\omega) \chi_{\delta\beta}(\omega). \qquad (6)$$

In the eq. (6), to make notations shorter we expressed the XC kernel and other matrices (defined below) in the form of a $4 \times 4$ matrix whose rows (columns) are defined by pair of the following indices: $1 = \uparrow\uparrow$, $2 = \uparrow\downarrow$, $3 = \downarrow\uparrow$ and $4 = \downarrow\downarrow$. The other matrices in the eq. (6) are the Fourier transform of the correlation function (generalized susceptibility) $\chi_{\sigma\sigma'\bar{\sigma}\bar{\sigma}'}(t) = -\sum_{a,b}\langle \hat{T} c_\sigma^{a+}(t) c_{\sigma'}^a(t) c_{\bar{\sigma}}^{b+}(0) c_{\bar{\sigma}'}^b(0) \rangle$ ($\chi^{(0)}_{\sigma\sigma'\bar{\sigma}\bar{\sigma}'}(\omega)$ in the non-interacting case) where $c_\sigma^a$ and $c_\sigma^{a+}$ are the electron annihilation and creation operators, respectively, a, b are the band indices, and $\hat{T}$ is the time-ordering operator, for details see Supplementary Information, Section I).

After calculation of the susceptibility with the DMFT approximation substitution of the result into eq. (6) and inversion the matrix equation give the following XC kernel matrix:

$$\hat{f}_{XC}^{DMFT}(\omega) = \begin{pmatrix} f_{XC11}(\omega) & 0 & 0 & 0 \\ 0 & 0 & f_{XC23}(\omega) & 0 \\ 0 & f_{XC32}(\omega) & 0 & 0 \\ 0 & 0 & 0 & f_{XC44}(\omega) \end{pmatrix}. \tag{7}$$

The time domain transformation of matrix in eq. (7) is substituted in eq. (3) to get XC potential. The non-diagonal elements of eq. (7) take spin flipping processes into account: for example, $f_{XC32}(\omega)$ accounts for the spin-up to spin-down transition. Finally, the limiting case of "no-memory" solution of the problem is obtained by approximating the XC kernel in eq. (7) as $\hat{f}_{XC}^{DMFT}(\omega) = \hat{f}_{XC}^{DMFT}(0)$ which in the real time representation becomes

$$f_{XC\sigma\sigma'\bar{\sigma}\bar{\sigma}'}(r, r', t, t') = \delta(r - r') f_{XC\sigma\sigma'\bar{\sigma}\bar{\sigma}'}^{DMFT}(0) \delta(t - t'). \tag{8}$$

The Kohn-Sham equation is (eq. (1)) is solved by using the Density-Matrix formalism in which the time-dependent density matrix elements $\rho_{k\sigma\sigma'}^{ln}(t)$ are calculated by propagating the Liouville equation

$$i\frac{\partial \rho^{ln}_{k\sigma\sigma'}(t)}{\partial t} = [H,\rho]^{ln}_{k\sigma\sigma'}(t) \equiv \sum_{s,\sigma''}\left(H^{ls}_{k\sigma\sigma''}(t)\rho^{sn}_{k\sigma''\sigma'}(t) - \rho^{ls}_{k\sigma\sigma''}(t)H^{sn}_{k\sigma''\sigma'}(t)\right), \quad (9)$$

where

$$H^{nl}_{k\sigma\sigma'}(t) = \int \psi^{n*}_{k\sigma}(r)\widehat{H}_{\sigma\sigma'}(r,t)\psi^{l}_{k\sigma'}(r)dr \quad (10)$$

are the time independent orbital-spin matrix elements of the Hamiltonian $\widehat{H}_{\sigma\sigma'}(r,t)$ given by

$$\widehat{H}_{\sigma\sigma'}(r,t) = \varepsilon^{n}_{k}\delta^{nl}\delta_{\sigma\sigma'} + \sum_{q<k_F,ab,\bar{\sigma}\bar{\sigma}'}\int_{-\infty}^{t}dt' F^{nlab}_{kq;\sigma\sigma'\bar{\sigma}\bar{\sigma}'}(t,t')\left(\rho^{ab}_{q\bar{\sigma}\bar{\sigma}'}(t') - \rho^{ab}_{q\bar{\sigma}\bar{\sigma}'}(0)\right) +$$

$$+ \vec{d}^{nl}_{k\sigma\sigma'} \cdot \vec{E}(t) \quad (11)$$

where matrix elements

$$F^{nlab}_{kq;\sigma\sigma'\bar{\sigma}\bar{\sigma}'}(t,t')$$

$$= \int \psi^{n*}_{k\sigma}(r)\psi^{l}_{\sigma'k}(r)\left[\delta_{\sigma\sigma'}\delta_{\bar{\sigma}\bar{\sigma}'}\frac{1}{|r-r'|} + f_{XC\sigma\sigma'\bar{\sigma}\bar{\sigma}'}(r,t,r',t')\right]\psi^{a}_{q\bar{\sigma}}(r')\psi^{b*}_{q\bar{\sigma}'}(r')drdr' \quad (12)$$

describe the correlation effects and

$$\vec{d}^{nl}_{k\sigma\sigma'} = e\int \psi^{m*}_{k\sigma}(r)\vec{r}(r,t)\psi^{l}_{k\sigma'}(r)dr \quad (13)$$

are the matrix elements of the dipole moment. In eq. (11), $\varepsilon^{n}_{k}$ are the static DFT eigen-energies, $\rho^{ln}_{k\sigma\sigma'}(t=0) = 0$ as initial occupancy of unoccupied site of band beyond Fermi energy whose occupancy is tracked over time and $E(t) = E_0 e^{-\frac{t^2}{\tau^2}}\cos(\omega t)$ is taken isotropic in space.

(see Supplementary Information, Section II for details).

One more remark – on the angular momentum dynamics of the system – has to be made. Earlier models assumed possible change of the angular momentum of the system due to photon absorption [15]. More popular (see, e.g., ref.[12,40,41]) is the scenario in which the spin-flip processes are accompanied by transfer of angular momentum to the lattice as a result of electron-phonon scattering. During the after-pulse dynamics, in the absence of electron-phonon scattering, non-collinear coupling can be sufficient to redistribute the angular momentum between the spin and the orbital momentum "subsystems". In absence of the spin-orbit interaction, one can expect ultrafast demagnetization driven by electron-electron correlation effects. Indeed, the electron-electron correlation effects may lead to a redistribution of the excited electrons from the majority to minority bands [42]. In this work, we consider the electron-electron correlation induced (non-collinear) scenario of the ultrafast demagnetization.

# Supplementary Information

## I. DMFT XC KERNEL MATRIX

In this brief Section, we give the general form of the four-index equation that connects the DMFT XC kernel matrix $f^{DMFT}_{XC\sigma\sigma'\bar{\sigma}\bar{\sigma}'}(\omega)$ and Fourier transform of the correlation function (generalized susceptibility) $\chi_{\sigma\sigma'\bar{\sigma}\bar{\sigma}'}(t) = -\sum_{a,b}\langle \hat{T} c^{a+}_{\sigma}(t) c^{a}_{\sigma'}(t) c^{b+}_{\bar{\sigma}}(0) c^{b}_{\bar{\sigma}'}(0)\rangle$:

$$\chi_{\sigma\sigma'\bar{\sigma}\bar{\sigma}'}(\omega) = \chi^{(0)}_{\sigma\sigma'\bar{\sigma}\bar{\sigma}'}(\omega) + \sum_{\sigma'',\sigma''',\bar{\sigma}'',\bar{\sigma}'''} \chi^{(0)}_{\sigma\sigma'\sigma''\sigma'''}(\omega) f^{DMFT}_{XC\sigma''\sigma'''\ \bar{\sigma}''\bar{\sigma}'''}(\omega) \chi_{\bar{\sigma}''\bar{\sigma}'''\bar{\sigma}\bar{\sigma}'}(\omega). \quad (\text{I.1})$$

As it is described in the main text, after introduction of the generalized two-spin indices $1 = \uparrow\uparrow$, $2 = \uparrow\downarrow, 3 = \downarrow\uparrow$ and $4 = \downarrow\downarrow$, the last equation can be transformed into $4 \times 4$ matrix eq. (6).

### II. DENSITY MATRIX TDDFT

In the density-matrix formalism, one expands the spin wave functions in terms of the static DFT Kohn-Sham occupied (v) and unoccupied (c) wave functions $\psi^l_{k\sigma}(r)$ as

$$\Psi_{k\sigma}(r,t) = \sum_{l \in v,c} c^l_{k\sigma}(t)\psi^l_{k\sigma}(r). \qquad (II.1)$$

For convenience, instead of the time-dependent coefficients $c^l_{k\sigma}(t)$, where k and l refer to wave vector and an index referring to the five d orbitals of Ni which are taken as basis to expand the wave function, respectively, one can consider the density matrix defined as

$$\rho^{ln}_{k\sigma\sigma'}(t) = c^l_{k\sigma}(t)c^{n*}_{k\sigma'}(t), \qquad (II.2)$$

whose diagonal and off-diagonal elements are the state occupancies and the probability of transition (polarization), respectively. The dynamics of magnetization change in the system (dm(t)) defined as

$$dm(t) = \int [n_{\uparrow\uparrow}(r,t) - n_{\downarrow\downarrow}(r,t)]d^3r - \int [n_{\uparrow\uparrow}(r,0) - n_{\downarrow\downarrow}(r,0)]d^3r,$$

can be obtained using the density-matrix elements as

$$dm(t) = \sum_{l=v, k<k_F} [\rho^{ll}_{k\uparrow\uparrow}(t) - \rho^{ll}_{k\downarrow\downarrow}(t)] - \sum_{l=v, k<k_F} [\rho^{ll}_{k\uparrow\uparrow}(0) - \rho^{ll}_{k\downarrow\downarrow}(0)]. \qquad (II.3)$$

The time-dependent density matrix elements are calculated by propagating the Liouville equation that they satisfy:

$$i\frac{\partial \rho^{ln}_{k\sigma\sigma'}(t)}{\partial t} = [H, \rho]^{ln}_{k\sigma\sigma'}(t) \equiv \sum_{s,\sigma''}\left(H^{ls}_{k\sigma\sigma''}(t)\rho^{sn}_{k\sigma''\sigma'}(t) - \rho^{ls}_{k\sigma\sigma''}(t)H^{sn}_{k\sigma''\sigma'}(t)\right), \qquad (II.4)$$

where

$$H^{nl}_{k\sigma\sigma'}(t) = \int \psi^{n*}_{k\sigma}(r)\widehat{H}_{\sigma\sigma'}(r,t)\psi^l_{k\sigma'}(r)dr, \qquad (II.5)$$

are the time independent orbital-spin matrix elements of the Hamiltonian $\widehat{H}_{\sigma\sigma'}(r,t)$ which is

$$\widehat{H}_{\sigma\sigma'}(r,t) = \varepsilon_k^n \delta^{nl}\delta_{\sigma\sigma'} + \sum_{q<k_F,ab,\bar\sigma\bar\sigma'} \int_{-\infty}^{t} dt' F_{kq;\sigma\sigma'\bar\sigma\bar\sigma'}^{nlab}(t,t') \left(\rho_{q\bar\sigma\bar\sigma'}^{ab}(t') - \rho_{q\bar\sigma\bar\sigma'}^{ab}(0)\right) +$$

$$\vec{d}_{k\sigma\sigma'}^{nl} \cdot \vec{E}(t) \tag{II.6}$$

where

$$F_{kq;\sigma\sigma'\bar\sigma\bar\sigma'}^{nlab}(t,t')$$

$$= \int \psi_{k\sigma}^{n*}(r)\psi_{\sigma'k}^{l}(r)\left[\delta_{\sigma\sigma'}\delta_{\bar\sigma\bar\sigma'}\frac{1}{|r-r'|} + f_{XC\sigma\sigma'\bar\sigma\bar\sigma'}(r,t,r',t')\right]\psi_{q\bar\sigma}^{a}(r')\psi_{q\bar\sigma'}^{b*}(r')dr dr' \tag{II.7}$$

and

$$\vec{d}_{k\sigma\sigma'}^{nl} = e \int \psi_{k\sigma}^{m*}(r)\vec{r}(r,t)\psi_{k\sigma'}^{l}(r)dr. \tag{II.8}$$

is the matrix elements of the dipole moment, $\varepsilon_k^n$ are the static DFT eigen-energies, $\rho_{k\sigma\sigma'}^{ln}(t=0) = 0$ as initial occupancy of unoccupied site of band beyond Fermi energy whose occupancy is tracked over time and $E(t) = E_0 e^{-\frac{t^2}{\tau^2}} \cos(\omega t)$ is taken isotropic in space.

In calculation, the time dependence of the nuclear motion is not considered since such a motion is negligible in the short time scale of this study and the time-dependence of the electron-electron Hartree interaction is not included since its contribution is small. XC term in interaction matrix elements (second term in eq. II.6) is approximated as

$$F_{kq;\sigma\sigma'\bar\sigma\bar\sigma'}^{nlab}(t,t') \approx A f_{XC\sigma\sigma'\bar\sigma\bar\sigma'}^{DMFT}(t-t'), \tag{II.9}$$

where $A$ is the orbital and spin averaged time-independent part:

$$A = \overline{\left(\int \psi_{k\sigma}^{n*}(r)\psi_{\sigma'k}^{l}(r)\psi_{q\bar\sigma}^{a}(r)\psi_{q\bar\sigma'}^{b*}(r)dr\right)}. \tag{II.10}$$

In eq. (II.6), the sum over all momenta q≤ $k_F$ is replaced by integration over energy $\varepsilon$ of occupied part of band for continuous representation, then the Hamiltonian in eq. (II.7) becomes

$$H^{nl}_{k\sigma\sigma'} = \varepsilon^n_k \delta^{nl} \delta_{\sigma\sigma'} + A \sum_{p\epsilon occ.\ q\epsilon unocc.,\bar{\sigma}\bar{\sigma}'} \int_{-\infty}^{t} dt'\ f^{DMFT}_{XC\sigma\sigma'\bar{\sigma}\bar{\sigma}'}(t-t') \left( \rho^{ab}_{q\bar{\sigma}\bar{\sigma}'}(t') - \rho^{ab}_{q\bar{\sigma}\bar{\sigma}'}(0) \right) \int \int d\varepsilon' d\varepsilon\ A^p(\varepsilon) A^q(\varepsilon') + \vec{d}^{nl}_{k\sigma\sigma'} \cdot \vec{E}(t) \qquad (II.11)$$

where occ. and unocc.in the summation sign refers to the occupied and unoccupied band and $A^p(\varepsilon)$ refers the DFT projected density of states. The presence of $f^{DMFT}_{XC\sigma\sigma'\bar{\sigma}\bar{\sigma}'}$ from DMFT with spin-flipping component, the non-collinear spin theory allows spin to flip.

The spin-flip transitions take place between the states obtained within DFT+DMFT, i.e. with a rather accurate (basically, ab initio) solution that takes into account effects of strong electron-electron correlations. Namely, the excited states that correspond to the spin-flip transitions are defined by poles of the charge susceptibility that gives the XC kernel. These excited states are the key element of the theory. They define the XC kernel and hence the time-dependent demagnetization. The key role of spin-flip excitations is also confirmed also by our calculations with approximate TDDFT solution, where the XC kernel is spin-independent, i.e. approximated by an average-over-spin function (the last function is still defined by the excitation energies of the system).

Moreover, the TDDFT+DMFT theory takes into account memory effects that are rather important at the fs time scale (at times of the e-e scattering) and are neglected in previous studies [1,2].

### III. FIELD-DEPENDECE OF THE DEMAGNETIZATION

The results for the dependence of the demagnetization on the pulse field amplitude are shown in Fig. SI.1. It is important to note that at large fields (the bottom curve in Fig. SI. 1) the value of the final demagnetization exceeds the initial magnetization of $0.64\ \mu_B$. We show the results for such strong fields in order to demonstrate the limits of applicability of the linear-response TDDFT for Ni, i.e. at what values of the pulse energy the approximation fails.

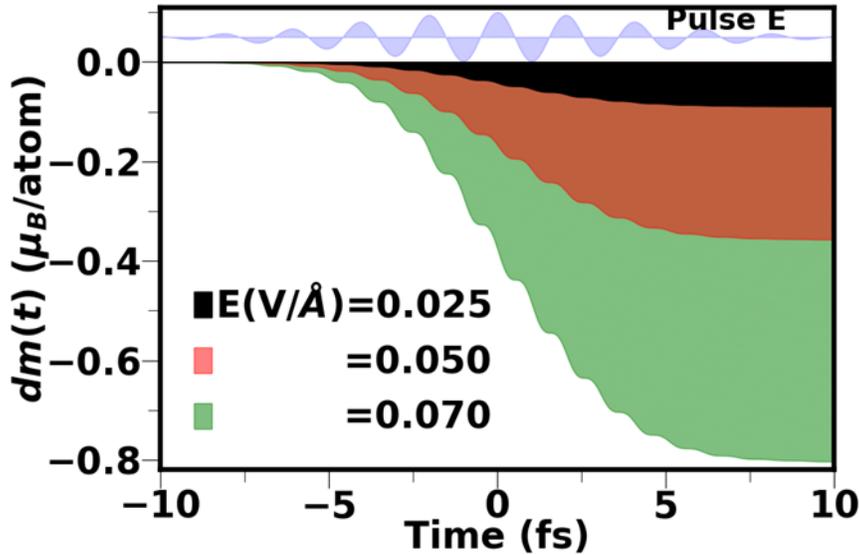

Fig. SI.1. The time-dependence of the TDDFT+DMFT demagnetization at different field amplitudes for pulse with ℏω = 2 eV and duration 7.2 fs.

## IV. ANGULAR MOMENTUM-DEPENDENCE OF THE DEMAGNETIZATION

The results for the dependence of the demagnetization on the z-components of the angular momentum are shown in Fig. SI.2. As it follows from our calculations, the m=-1 and m=1 components experience equal change, also the amount of change of the m=-2 and m=2 components is rather similar. This results in rather small change of the total component of the angular momentum (green area).

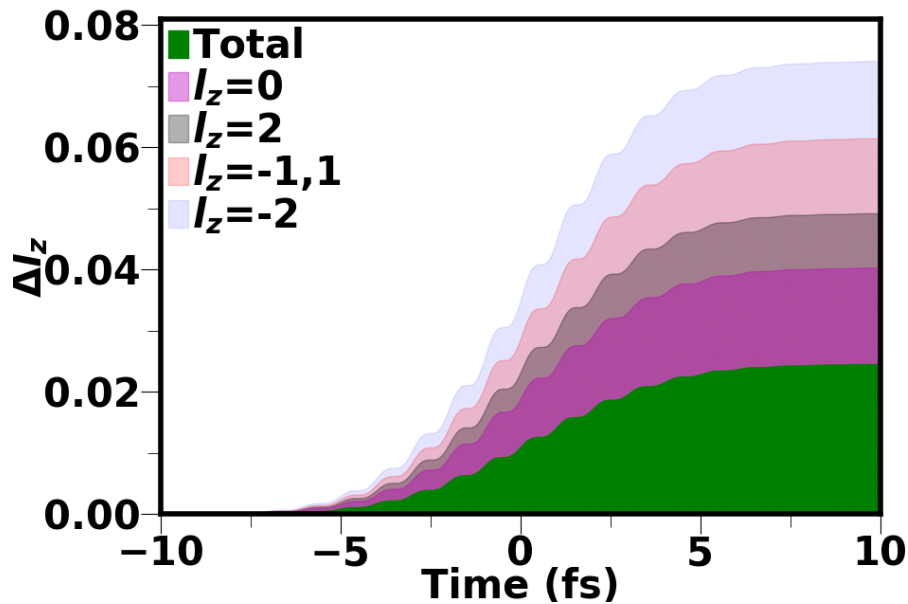

Figure SI.2. Change of the total z-component of the angular momentum and of its different components due to laser pulse perturbation (the pulse parameters are the same as in Fig. 4).